# Structure-aware Registration Network for Liver DCE-CT Images

Peng Xue, Jingyang Zhang, Mianxin Liu, Yuning Gu, Jiawei Huang, Feihong Liu, Yongsheng Pan, Lei Ma, Xiaohuan Cao, Dinggang Shen *Fellow, IEEE*

**Abstract**—Image registration of liver dynamic contrast-enhanced computed tomography (DCE-CT) is crucial for diagnosis and image-guided surgical planning of liver cancer. However, intensity variations due to the flow of contrast agents combined with complex spatial motion induced by respiration brings great challenge to existing intensity-based registration methods. To address these problems, we propose a novel structure-aware registration method by incorporating structural information of related organs with segmentation-guided deep registration network. Existing segmentation-guided registration methods only focus on volumetric registration inside the paired organ segmentations, ignoring the inherent attributes of their anatomical structures. In addition, such paired organ segmentations are not always available in DCE-CT images due to the flow of contrast agents. Different from existing segmentation-guided registration methods, our proposed method extracts structural information in hierarchical geometric perspectives of line and surface. Then, according to the extracted structural information, structure-aware constraints are constructed and imposed on the forward and backward deformation field simultaneously. In this way, all available organ segmentations, including unpaired ones, can be fully utilized to avoid the side effect of contrast agent and preserve the topology of organs during registration. Extensive experiments on an in-house liver DCE-CT dataset and a public LiTS dataset show that our proposed method can achieve higher registration accuracy and preserve anatomical structure more effectively than state-of-the-art methods.

*Index Terms*—Image registration, liver DCE-CT, segmentation-guided registration network, structure-aware constraints

## I. INTRODUCTION

L IVER dynamic contrast-enhanced computed tomography (DCE-CT) imaging typically acquires a sequence of images before and after injection of contrast agent, namely pre- contrast phase, arterial phase and venous phase (both called post-contrast phase). Precise registration between DCE-CT images at pre- and post-contrast phases can obtain refined subtraction images to reveal the flow of contrast agent and expose the focuses (e.g. tumors), which is vital for quantitative analysis of diagnosis and therapy assessment [1], [2]. For liver cancer, surgery is an effective treatment manner. DCE-CT can provide complementary information from different phases, particularly the arterial and venous phases. The two phases can provide different information of tumors and their surroundings, thus helping clinicians evaluate the tumor comprehensively and performing more accurate surgical planning. In this work, we focus on one exemplar application of liver DCE-CT image registration which is to register images of arterial phase with venous phase for assisting the design of surgical planning.

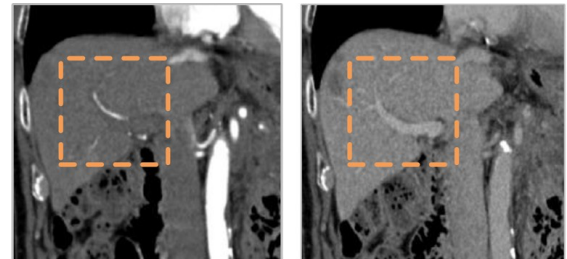

(a) Arterial phase　　　　(b) Venous phase

Fig. 1. Liver DCE-CT images at arterial phase and venous phase. Hepatic artery and portal vein are enhanced at the arterial phase and venous phase, respectively.

In general, there are two major challenges for liver DCE-CT image registration: 1) Large motion of subtle anatomies within the liver. A DCE-CT scan typically requires a few minutes, during which the motion of patient, including respiratory motion, heart motion, and stomach and bowel peristalsis, is inevitable. The large motion makes it difficult to align subtle anatomies accurately, especially for vessels including both arteries and veins. 2) Intensity or local appearance variations due to the uptake and washout of contrast agent. As shown in Fig. 1, due to the flow of contrast agent, the hepatic artery and the portal vein exhibit similar intensity values at the arterial and the venous phase, respectively. Although the two types of vessels are adjacent spatially, the arteries and veins should not be overlapped when performing registration between these two phases. However, for such scenario, most intensity-based

This work was supported in part by National Natural Science Foundation of China under Grant 62131015, Science and Technology Commission of Shanghai Municipality (STCSM) under Grant 21010502600, and The Key R&D Program of Guangdong Province, China under Grant 2021B0101420006. *(Peng Xue and Jingyang Zhang contributed equally to this work.) (Corresponding authors: Xiaohuan Cao, Dinggang Shen.)*

Peng Xue, Jingyang Zhang, Yuning Gu, Jiawei Huang, Feihong Liu, Yongsheng Pan, Lei Ma are with School of Biomedical Engineering, ShanghaiTech University, Shanghai, China.

Xiaohuan Cao is with Shanghai United Imaging Intelligence Co., Ltd.,Shanghai, China (e-mail: xiaohuan.cao@uii-ai.com).

Dinggang Shen is with School of Biomedical Engineering, Shang- haiTech University, Shanghai 201210, China. He is also with Shanghai United Imaging Intelligence Co., Ltd., Shanghai 200230, China, and Shanghai Clinical Research and Trial Center, Shanghai 201210, China (email: Dinggang.Shen@gmail.com).

registration methods may easily align these two different types of vessels incorrectly.

For tackling the large motion of subtle anatomies, a general way is to introduce a novel regularization term, such as isotropic total variation regularization [3], topology preservation term [4] or hyper-elastic regularization [5], into energy function to preserve the topological structure of the entire deformation field. Although these methods can partially improve the registration performance, they usually regularize the global deformation field in a uniform manner, which remains insufficient to process the images with large local deformations.

In order to handle the intensity variations between the fixed and moving images, several methods employ intensity-insensitive metrics, such as normalized gradient fields [6], Lorentzian estimator [7], and modality-independent neighborhood descriptor [8] to guide the registration. However, these similarity metrics cannot capture underlying anatomical information of organs to guarantee reasonable registration results.

Another way to reduce the effect of contrast variation in registration is to separate the motion components from the intensity variations caused by contrast agents. If enhanced components can be separated from post-contrast images, the remaining de-enhanced images can be easily registered by conventional intensity-based registration methods. Following this strategy, registration methods based on statistical models such as principal component analysis (PCA) [9], independent component analysis (ICA) [10], and correlation-weighted sparse representation [11] are proposed. Although the above de-enhancing methods [9]–[11] have shown relatively promising performance, the limited image samples from DCE-CT sequence is insufficient for effectively performing PCA or ICA.

Furthermore, segmentation-guided registration strategy [12]–[14] is another way to separate contrast-enhanced regions from post-contrast images. Specifically, corresponding structures can be segmented reliably with anatomical knowledge. Such segmentation results can then be used to highlight the organs of interest, focused regions, or other anatomical structures between paired images, and serve as guidance for alignment of voxels. Therefore, the segmentation-guided registration strategy could avoid the influence of intensity variations effectively, and it has been widely used in various registration tasks.

However, some limitations in existing segmentation-guided registration methods need to be further investigated for their application to liver DCE-CT images. First, existing methods only focus on volumetric registration inside the organ masks, ignoring the inherent geometric structure of the organs (e.g., the organ surface, organ morphology, tubular structures of vessels). More importantly, most segmentation-guided registration methods [12]–[14] require paired segmentations for guidance. In some special scenarios, such as liver DCE-CT images, it can only obtain segmentations of either arterial or venous vessels at different phases (as shown in Fig. 1), and existing registration methods are incapable of utilizing such unpaired segmentations.

For tackling the above issues in liver DCE-CT image registration, a structure-aware registration network is proposed in this paper. The structure-aware constraints are designed based on the geometric information of different organs from hierarchical perspectives. In addition, we propose a non-overlap loss to utilize the unpaired segmentations of blood vessels in arterial phase and venous phase. Moreover, the designed structure-aware constraints are imposed on forward and backward registrations bidirectionally to relieve the bias in the guidance from unpaired segmentations.

Specifically, there are three highlights of our work:

1) We explore the structural characteristics of various organs from multiple geometrically meaningful perspectives (e.g., centerline, surface and volume of vessel). Then, a novel structure-aware and segmentation-guided deep registration network is designed by leveraging structural information to handle the effect of large local motion, and varied contrast-enhancement between fixed and moving images.

2) We introduce several novel structure-aware constraints on deformation field. Different from existing global regularization constraints, our proposed structure-aware constraints fully consider a dense relationship of displacement within the extracted structure, and can preserve their topology of segmented organs from hierarchical perspectives of centerline and surface.

3) In order to fully utilize all available segmentations for registration, we employ volumetric overlap and non-overlap losses for paired segmentations and unpaired segmentations, respectively. Meanwhile, we impose the structure-aware constraints on forward and backward deformation fields simultaneously to prevent bias introduced by the unpaired segmentations, and improve the registration performance effectively.

## II. RELATED WORK

### A. Conventional Image Registration Methods

Conventional image registration methods, such as elastic [15], fluid [16], B-spline [17] or Markov random field (MRF) model [4], are usually based on the iterative optimization. Although the conventional image registration methods can solve a variety of registration tasks, there still remains some challenging problems that have not been well solved. For instance, it is difficult to accurately register images with large local deformations while keeping the smoothness of deformation field. Furthermore, the multi-modal image registration, or registration of images with domain shift, needs to be further explored. More importantly, the conventional optimization-based methods are computationally complex, and the efficiency cannot sufficiently meet the clinical requirements.

### B. Deep Learning-based Image Registration Methods

Compared with conventional registration methods, deep learning-based methods can significantly improve the registration speed while achieving comparable performance at the same time. The convolutional neural network is designed to learn a complex mapping model between the to-be-registered image pair. Based on the network design and data information used to train the registration model, the deep learning-based registration can be summarized as three categories:

(1) Supervised learning that uses the ground-truth deformations as the label to train the model [18]–[20]. For supervised registration network, the training loss is usually defined by the distance metric between the predicted and the ground-truth deformation fields. Although the results of such methods are encouraging, the performance is affected by the quality of the generated ground truth and also the size of training data.

(2) Unsupervised learning that mainly uses image similarity as the guidance to train the model without ground-truth labels. Such method is preferred for registration task since it does not need the "ground-truth" transformations. Basically, the definition of loss function has two main terms: the image similarity after registration and the smoothness (regularization) of the predicted deformation field. A representative unsupervised network is VoxelMorph [21], [22], which uses variational methods to predict deformation fields and maintain diffeomorphism of deformation fields by integrating velocity fields. Due to the fact that it does not require extensive data annotation, its registration accuracy is not limited by the quality of the ground truth, and it can be generalized to a wide variety of clinical applications.

(3) Weakly-supervised learning where part of segmentations are applied during the model training. The weakly-supervise learning can be understood as training the registration model using some auxiliary information such as organ segmentations as the guidance to train the registration network for higher accuracy. For instance, Hering et al. [23] used both weak and dual supervision by incorporating segmentation similarity and image similarity to the loss function for cardiac motion tracking. In addition, Hu et al. [24] proposed an end-to- end network to predict both affine transformation and local deformations, and the network is trained by the overlap rate of organ segmentations. By introducing auxiliary information, these weakly-supervised methods can improve registration performance compared with pure unsupervised training strategies.

However, existing weakly-supervised registration network still have some limitations. First, they do not capture the inherent topological structure of organs during registration, leading to disconnection of small vessels and non-smoothness of organ surfaces. Second, limited by the image quality and physical acquisition process, paired organ segmentations of moving and fixed images are not always available in medical imaging, hampering clinical application of segmentation-guided registration method.

## III. METHOD

### A. Naive Weakly-Supervised Deep Registration Network

The goal of deformable image registration aims to establish spatial correspondence between $n$-dimensional fixed image $I_F : \Omega \in \mathbb{R}^n$ and $n$-dimensional moving image $I_M : \Omega \in \mathbb{R}^n$. It aims to find the deformation $\varphi(p)$ for each voxel $p \in \Omega$ such that $I_F(p)$ and $[I_M \circ \varphi](p)$ correspond to similar anatomical location, where $\varphi = Id + \mathbf{u}$ is the deformation field characterized by a displacement field $\mathbf{u}$ and $Id$ represents the identity transform [25]. Existing deep learning-based registration framework usually applies an encoder-decoder network (referred as U-net or V-net) to estimate the deformation $\varphi$ from a pair of inputted images ($I_F$ and $I_M$).

When using an unsupervised learning problem setting, the registration network minimizes a loss function $\mathcal{L}$ that measures the dissimilarity between the fixed image $I_F$ and the warped moving image $I_M \circ \varphi$ :

$$\mathcal{L} = \mathcal{L}_{\text{sim}}(I_F, I_M \circ \varphi) + \lambda \mathcal{L}_{\text{reg}}(\varphi) \qquad (1)$$

where $\mathcal{L}_{\text{sim}}(I_F, I_M \circ \varphi)$ represents the dissimilarity loss, which is used to enforce $I_M \circ \varphi$ to be similar to $I_F$. $\mathcal{L}_{\text{reg}}(\varphi)$ is a regularization term to preserve smoothness of the predicted deformation field $\varphi$.

Following the typical weakly-supervised registration manner, several methods [23], [24] leverage the auxiliary information, such as organ segmentations, to train the registration network. If a deformation field $\varphi$ represent accurate anatomical correspondence, the regions in $I_F$ and $I_M \circ \varphi$ corresponding to the same organ or anatomical structure should overlap well. Adopting this strategy, an auxiliary loss can be defined using some classical overlap metrics, such as Dice score, Jaccard, and cross-entropy. For example, the volumetric overlap of same anatomical structure can be quantified using Dice score:

$$\text{Dice}(S_F, S_M \circ \varphi) = 2 \times \frac{|S_F \cap (S_M \circ \varphi)|}{|S_F| + |S_M \circ \varphi|} \qquad (2)$$

where $S_F$ and $S_M \circ \varphi$ represent the voxels of same anatomical structure for $I_F$ and $I_M \circ \varphi$, respectively. Then, an auxiliary loss based on segmentations can be defined as:

$$\mathcal{L}_{\text{seg}}(S_F, S_M \circ \varphi) = -\text{Dice}(S_F, S_M \circ \varphi) \qquad (3)$$

With the help of segmentation results, equation (3) have been proved to significantly improve the registration performance.

### B. Structure-aware Registration Network

In general, the morphological or topological structures of different organs are not identical. We use these structural information to supervise registration network, to estimate the deformation field more effectively. Generally, the structural information of organs can be effectively characterized from several key geometric perspectives. For instance, volumetric structures, such as liver parenchyma, bladder, and rectum, can be effectively represented as 3D surfaces or 2D contours. And tubular structures, such as blood vessels, can be modeled by skeleton lines.

Therefore, in this paper, for different anatomies, we utilize geometric representations for various organs and propose structure-aware constraints to preserve the organ morphology after registration. Specifically, the proposed structure-aware registration network is designed as shown in Fig. 2. The network takes pairs of images $I_F$ and $I_M$ with their segmentations of liver and blood vessels ($S_F^l$, $S_F^v$ and $S_M^l$, $S_M^v$) as inputs, computes the forward deformation $\varphi$ and inverse deformation $\varphi^{-1}$. Among them, $S_F^l$ and $S_M^l$ represent the paired liver segmentations, $S_F^v$ and $S_M^v$ represent different types of vessel segmentations in the venous and arterial phases, respectively. Then, we warp the moving image $I_M$ and its corresponding segmentation masks $S_M^l$ and $S_M^v$ to $I_M \circ \varphi$, $S_M^l \circ \varphi$, and $S_M^v \circ \varphi$, respectively, for the evaluations of the similarity between $I_M \circ \varphi$ and $I_F$, the overlap between $S_M^l \circ \varphi$ and $S_F^l$, and the non-overlap between $S_M^v \circ \varphi$ and $S_F^v$.

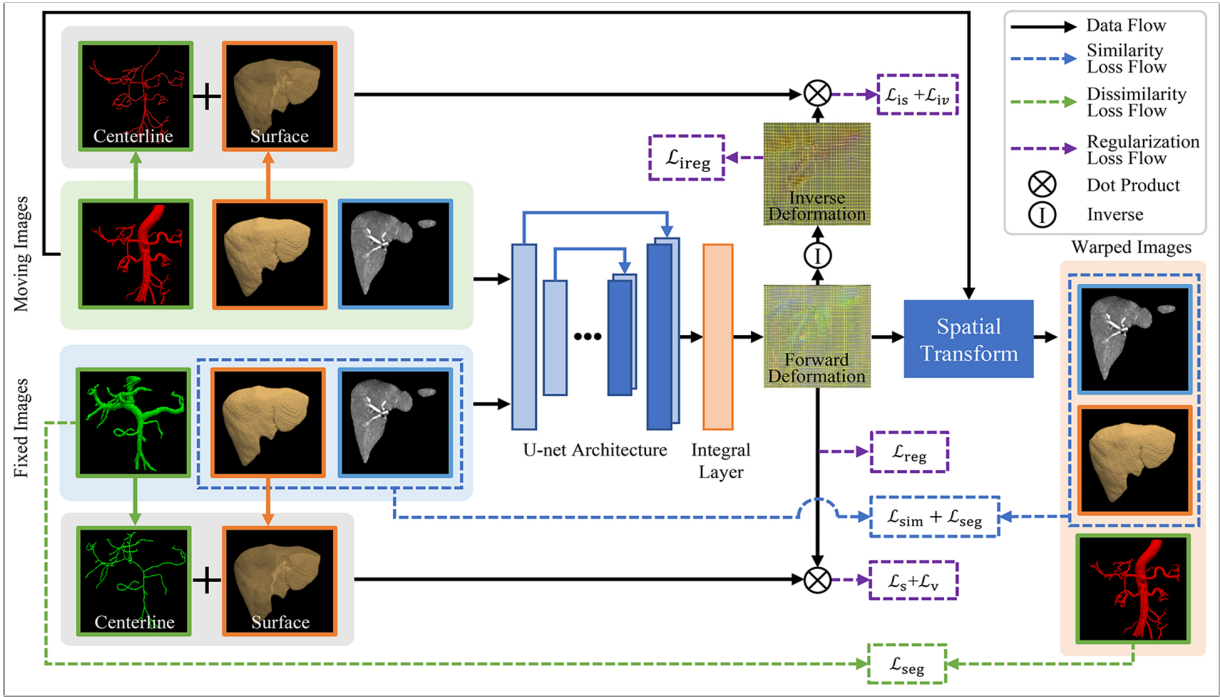

Fig. 2. An overview of our proposed structure-aware registration network. The structure information is extracted from by analyzing the structural characteristics of different types of organs based on segmentation maps. Then, structure-aware constraints are designed according to the extracted geometric information, and imposed on forward and backward deformation field simultaneously.

In addition, for arterial and venous vessels, we extract their centerlines $C_F^v$ and $C_M^v$ from $S_F^v$ and $S_M^v$, respectively. to represent their organ morphology. For liver parenchyma with large volume, paired surfaces $O_F^l$ and $O_M^l$ are extracted from $S_F^l$ and $S_M^l$ to represent the topology of the whole liver. Then, the structure-aware constraints that can preserve organ topologies are constructed using extracted structural information, and imposed on the forward deformation field $\varphi$ and inverse deformation field $\varphi^{-1}$ simultaneously to avoid the bias introduced by the unpaired segmentations.

In summary, there are two different types of loss functions for the entire structure-aware registration network: 1) the losses of similarity term between warped images ($I_M \circ \varphi$, $S_M^l \circ \varphi$, $S_M^v \circ \varphi$) and fixed images ($I_F$, $S_F^l$, $S_F^v$), and 2) the losses of local and global regularization terms for both forward deformation field $\varphi$ and inverse deformation field $\varphi^{-1}$.

$$\mathcal{L}_{all} = \mathcal{L}_{sim}(I_F, I_M \circ \varphi) + \lambda_1 \mathcal{L}_{seg}^l(S_F^l, S_M^l \circ \varphi)$$
$$+ \lambda_2 \mathcal{L}_{seg}^v(S_F^v, S_M^v \circ \varphi) + \lambda_3 \left( \mathcal{L}_{reg}(\varphi) + \mathcal{L}_{ireg}(\varphi^{-1}) \right)$$
$$+ \lambda_4 \left( \mathcal{L}_s(O_F^l, \varphi) + \mathcal{L}_{is}(O_M^l, \varphi^{-1}) \right)$$
$$+ \lambda_5 \left( \mathcal{L}_v(C_F^v, \varphi) + \mathcal{L}_{iv}(C_M^v, \varphi^{-1}) \right) \qquad (4)$$

where $\lambda_{i \in \{1,2,3,4,5\}}$ are hyper-parameters that can be empirically set to different values according to the experiment. $\mathcal{L}_{sim}(I_F, I_M \circ \varphi)$ captures the dissimilarity between $I_M \circ \varphi$ and $I_F$, and we normalize cross-correlation (NCC) to evaluate the dissimilarity between $I_M \circ \varphi$ and $I_F$. At the same time, as paired liver segmentation $S_F^l$ and $S_M^l$ are available, we compute the Dice loss in (3) as the volumetric overlap loss $\mathcal{L}_{seg}^l(S_F^l, S_M^l \circ \varphi)$ to ensure accurate alignment of the liver parenchyma. For unpaired blood vessel segmentations $S_F^v$ and $S_M^v$, we directly use Dice score in (2) as non-overlap measure $\mathcal{L}_{seg}^v(S_F^v, S_M^v \circ \varphi)$ to prevent overlapping between different types of vessels (i.e., arteries and veins). Based on this, we use squared $\ell$-1 norm derivatives of the displacement field $\mathbf{u}$ as global regularization term $\mathcal{L}_{reg}(\varphi)$ and $\mathcal{L}_{ireg}(\varphi^{-1})$ to preserve smoothness of the entire predicted deformation $\varphi$ and $\varphi^{-1}$.

However, only relying on the above-mentioned losses cannot ensure the preservation of the topology of organs during registration. Therefore, according to the morphology of different organs, we construct structure-aware constraints $\mathcal{L}_s(O_F^l, \varphi)$, $\mathcal{L}_{is}(O_M^l, \varphi^{-1})$, $\mathcal{L}_v(C_F^v, \varphi)$, $\mathcal{L}_{iv}(C_M^v, \varphi^{-1})$ to preserve the topology of organs during registration, which will be elaborated in the following subsection.

## C. Anatomical Structure Representation and Structure-aware Constraint

Generally, most existing registration methods constrain deformation fields to be globally smooth and continuous to avoid physically implausible displacement. For instance, the widely-used global smoothness regularization penalizes the squared $\ell$-1norm derivatives of the deformation field components:

$$\mathcal{L}_{\text{reg}}(\varphi) = \sum_{p \in \Omega} \|\nabla \mathbf{u}(p)\|^2 \tag{5}$$

where $\nabla \mathbf{u}(p) = (\frac{\partial \mathbf{u}(p)}{\partial x}, \frac{\partial \mathbf{u}(p)}{\partial y}, \frac{\partial \mathbf{u}(p)}{\partial z})$, $\frac{\partial \mathbf{u}(p)}{\partial x} \approx \mathbf{u}([p_x + 1, p_y, p_z]) - u([p_x, p_y, p_z])$, and similar for $\frac{\partial \mathbf{u}(p)}{\partial y}$ and $\frac{\partial \mathbf{u}(p)}{\partial z}$. However, a global homogeneous smoothing neglects local structural information of organs, and can lead to violation of the topology of organs during registration. Therefore, additional constraints should also be considered for registration.

For large volume like liver parenchyma, surface can well present the organ morphology. Thus, the smoothness of surface is crucial to preserve the organ topology. To keep the liver surface smooth after registration, the deformation of adjacent voxels should be locally consistent. The detailed implementation is shown in Fig. 3. Note that we use $O_M^l$ and forward deformation field $\varphi$ to illustrate the construction of $\mathcal{L}_s(O_F^l, \varphi)$, and the construction of $\mathcal{L}_{is}(O_M^l, \varphi^{-1})$ is similar to $\mathcal{L}_s(O_F^l, \varphi)$. First, the surface of liver $O_F^l \in \Omega$ is reconstructed from liver segmentation result by edge extraction algorithm (such as Canny operator [26] or Sobel operator [27]). Then, a structure-aware constraint is constructed according to $O_F^l$ and deformation field $\varphi$.

Here, we take a 2D slice as an example to illustrate. In a 2D slice, each voxel $p^l$ has 8 neighboring voxels as shown in Fig. 3. Among them, the dark blue voxels $p^l \in O_F^l$ belonging to the liver surface $O_F^l$ should have similar deformations $\mathbf{u}(p^l)$ with the center voxel, so as to keep the smoothness of liver surface. However, equation (5) only calculates gradient of $\mathbf{u}(p^l)$ along the direction of coordinate axis (as colored in red and green lines), ignoring the relation of deformations along the diagonal (as colored in blue and orange lines). In this way, equation (5) only constrains voxels along vertical and horizontal directions to have similar displacements as the center voxel, which cannot constrain other voxels along diagonal direction.

Therefore, we design a novel regularization term $\mathcal{L}_s(O_F^l, \varphi)$, which considers 8 neighboring voxels not only along the direction of the coordinate axis, but also along the gradient (directional derivative) of the diagonal direction:

$$\mathcal{L}_s(O_F^l, \varphi) = \sum_{p^l \in O_F^l} \|\nabla' \mathbf{u}(p^l)\|^2 \tag{6}$$

where $\nabla' \mathbf{u}(p^l) = (\frac{\partial \mathbf{u}(p^l)}{\partial x}, \frac{\partial \mathbf{u}(p^l)}{\partial y}, \frac{\partial \mathbf{u}(p^l)}{\partial m}, \frac{\partial \mathbf{u}(p^l)}{\partial n})$. Similar to forward differences $\frac{\partial \mathbf{u}(p^l)}{\partial x}$ and $\frac{\partial \mathbf{u}(p^l)}{\partial y}$ along directions of the $x$ and $y$ axes,

$$\frac{\partial \mathbf{u}(p^l)}{\partial m} \approx \frac{u([p_x^l + 1, p_y^l + 1]) - u([p_x^l, p_y^l])}{\sqrt{v_x^2 + v_x^2}} \tag{7}$$

and

$$\frac{\partial \mathbf{u}(p^l)}{\partial n} \approx \frac{u([p_x^l - 1, p_y^l + 1]) - u([p_x^l, p_y^l])}{\sqrt{v_x^2 + v_x^2}} \tag{8}$$

represent the forward differences along directions of $m$ (direction along blue line) and $n$ (direction along orange line), and $v_x$ and $v_y$ represent spatial resolutions along the $x$ and $y$

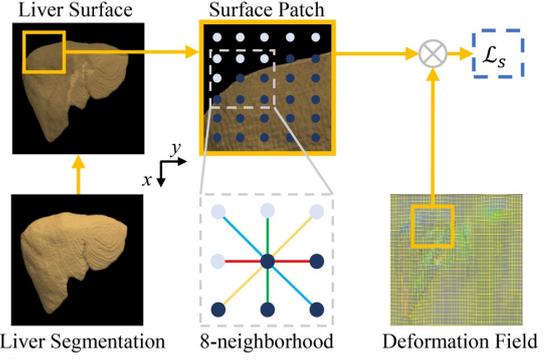
Fig. 3. The construction of structure-aware constraints for segmentation of liver.

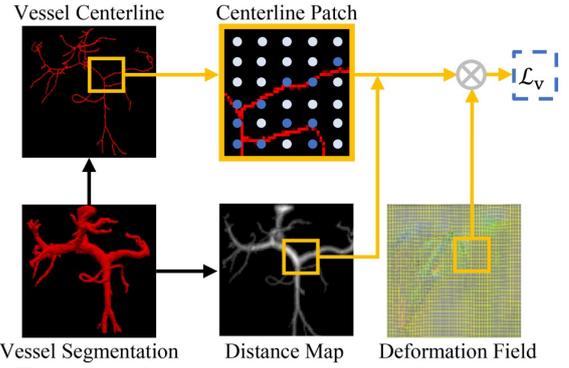
Fig. 4. The construction of structure-aware constraints for segmentation of vessel.

axes.

For liver vessels with tubular structures, the centerline can better represent their morphology. In order to maintain the morphology of vessels after registration, the deformation of adjacent voxels within the centerline should also be locally consistent. Similarly, we use $C_F^v$ and forward deformation field $\varphi$ to illustrate the construction of $\mathcal{L}_v(C_F^v, \varphi)$. As shown in Fig. 4, centerlines $C_F^v \in S_F^v$ of vessels are extracted from vessel segmentation $S_F^v$ using a classical skeleton extraction algorithm [28]. Then, the structure-aware constraints of vessels are constructed in a similar way to $\mathcal{L}_s(O_F^l, \varphi)$. It should be noted that the structure-aware constraints of vessels are constructed separately according to unpaired vessel segmentations, and imposed to the forward and backward deformation fields, respectively. In this way, unpaired segmentations can be fully exploited to preserve vessel morphology while avoiding bias at the same time.

In addition, to further preserve the topology of fine vessel branches, the structure-aware constraints are imposed with different weights along the centerline $C_F^v$. Specifically, a distance map $\text{dm}(p)$ is calculated based on vessel segmentation $S_F^v$. Among them, each voxel $p \in S_F^v$ measures the shortest distance between $p$ and background. Then, the structure-aware constraints $\mathcal{L}_v(C_F^v, \varphi)$ are constructed by applying different weights $w(p^v) = \frac{1}{\exp(\text{dm}(p^v))}, p^v \in C_F^v$. In this way, a large-weighted constraint is imposed on fine vessel branches to preserve the topology of tiny vessel branches. In summary, the structure-aware constraints on centerline can be formulated as:

$$\mathcal{L}_{\mathrm{v}}(C_F^v, \varphi) = w(p^v) \sum_{p^v \in C_F^v} \|\nabla' \mathbf{u}(p^v)\|^2$$

### D. Implementation

We have implemented the structure-aware network by Pytorch, and trained it on a single 20 GB NVIDIA V100 GPU. Adam optimizer is used with a learning rate of 1e-4. The structure-aware network is trained by 500 epochs with 200 iterations in each epoch. The five weights $\lambda_{i \in \{1,2,3,4,5\}}$ in (4) are set to 5, 4, 1, 0.5, and 1 empirically. Furthermore, we use a similar way as in [29] to generate the inverse deformation field $\varphi^{-1}$.

## IV. EXPERIMENTS AND RESULTS

In this section, we first conduct extensive intra-subject registration experiments on an in-house liver DCE-CT dataset, to evaluate our proposed method and compare it with state-of-the-art methods. In addition, we also conduct inter-subject experiments on a public LiTS dataset to further verify the robustness of our proposed method.

### A. Evaluation Metric

To evaluate the registration performance, we calculate the metrics of Dice similarity coefficient (DSC), relative absolute volume difference (RAVD), average symmetric surface distance (ASSD), and maximum symmetric surface distance (MSSD) be- tween the paired liver segmentations. In general, higher DSC, smaller RAVD, smaller ASSD, and smaller MSSD of liver segmentations represent better performance of the registration method.

Similarly, we also calculate the metric of DSC between unpaired vessel segmentations to evaluate the overlap between arteries and veins. Note that, the smaller DSC of unpaired vessel segmentations indicates less overlap and more reasonable registration results. In addition, in order to further evaluate morphological preservation of vessels, we calculate the number of connected regions for the warped vessel segmentation to evaluate disconnection of vessels.

### B. Evaluation on In-house Liver DCE-CT Dataset

#### 1) Dataset:
The DCE-CT dataset used in this paper is in-house data, with two-phasic (arterial phase and venous phase) 3D DCE-CT image sequence of 84 patients. Among them, the liver is segmented in both arterial and venous phases. For vessel segmentation, the arteries and veins are segmented from arterial phase and venous phases, respectively. All these segmentations are labelled by two medical imaging research fellows using an in-house voxel painting tool on the original image data. The resolution of the data is $0.70 \times 0.70 \times 0.87 mm^3$, and the image size is $353 \times 280 \times 230$. In order to avoid degradation of image quality caused by resampling, we directly register the images at their original resolution. In this study, we randomly split our DCE-CT dataset into 60 and 24 subjects for training and testing, respectively.

#### 2) Pre-Processing:
For liver DCE-CT image sequence, the intensities of corresponding blood vessels are inconsistent due to the flow of contrast agent. Therefore, a concise and effective image pre-processing workflow is critical for subsequent registration. The specific pre-processing workflow is shown in Fig. 5. First, for each subject, rigid alignment of all the images (including raw DCE image and corresponding segmentations) at arterial phase with the images at venous phase are performed. Then, we mask out non-liver features (including bones, kidneys, spleen, and stomach) to make our proposed network focus on the registration of liver regions. Subsequently, the intensities are normalized to 0-1, to highlight features within the liver. Afterwards, blood vessels and other characteristics can be easily identified from the processed image (as shown in Fig. 5). However, the intensity values of corresponding blood vessels are still inconsistent at different phases, which could make the registration performance unstable. To resolve this issue, a windowed median filter with window $3 \times 3 \times 3$ is first applied to each 3D image for noise suppression purpose. Then, the main blood vessels are enhanced using Frangi filter [30]. Note, the Frangi filer is only applied for the purpose of enhancing main vessels. After that, the enhanced blood vessels are superimposed on the filtered image, so that the corresponding vessels at different phases have similar intensity representation.

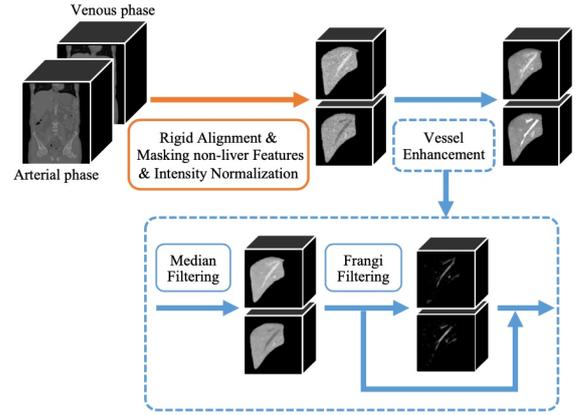

Fig. 5. The workflow of preprocessing for liver DCE-CT dataset.

#### 3) Comparative Analysis:
In order to comprehensively analyze the performance of the proposed method, we select 3 state-of-the-art registration methods (ANTs, NiftyReg and VoxelMorph) for comparative analysis. Among them, the deep learning-based VoxelMorph method can be used as a baseline to measure the validity of registration, and we train this model using its default parameters. In addition, two conventional registration methods (ANTs and NiftyReg) are also compared. We use Symmetric Normalization (SyN) implementation in the ANTs software package, with a mutual information (MI) similarity metric. In NiftyReg, we use B-spline based non-linear deformation with MI as the similarity metric and the stochastic gradient descent optimizer.

Table I lists the registration performance of various methods for the 24 testing subjects in the liver DCE-CT dataset, evaluated with paired liver segmentations and unpaired vessel segmentations. As shown in Table I, all compared methods can achieve relatively high DSC and small RAVD between paired liver segmentations. In addition, compared with other methods, the distance-based metrics, ASSD and MSSD, of our proposed methods are significantly ($p < 0.05$ in paired t-test) improved, which means that our proposed method exhibits better registration accuracy for the surface of liver. On the other hand, in terms of registration accuracy for vessels, our proposed method offers significant smaller DSC between unpaired vessel segmentations, and less vessel disconnection compared with other registration methods. In addition, we also calculate the computation time for all methods, including inference time for deep learning-based method using GPU and optimization time for traditional methods using CPU. It can be seen from Table I that the deep learning-based methods have huge boost of speed. Compared with VoxelMorph, the computation time of our proposed method is slightly longer, which is due to the fact that our proposed network has 6-channel input while the VoxelMorph has 2-channel input.

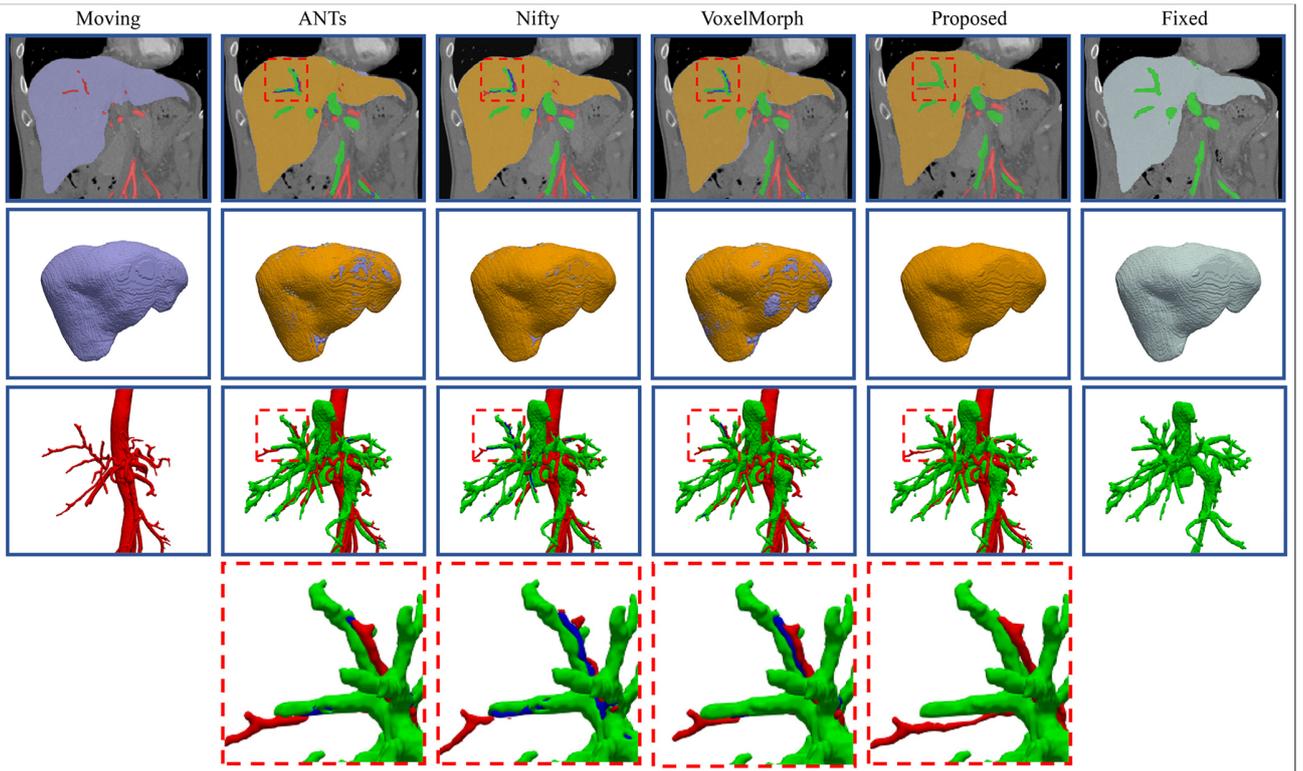

Fig. 6. The registration results of various methods in the liver DCE-CT dataset. The orange region represents the overlap between registered liver segmentation and fixed liver segmentation, and the blue region represents the overlap between arteries and veins. The vessels colored in red and green represent the hepatic artery and portal vein, respectively.

TABLE I
THE REGISTRATION PERFORMANCE OF VARIOUS METHODS IN LIVER DCE-CT DATASET, INCLUDING MEAN ± STD DSC, RAVD, ASSD, MSSD OVER LIVER SEGMENTATIONS, AND MEAN ± STD DSC, NUMBER OF BREAKS OVER REGISTERED VESSELS, AS WELL AS MEAN RFP FOR DEFORMATION FIELD AND COMPUTATION TIME FOR VARIOUS METHODS. THE BEST PERFORMANCE WITH A STATISTICAL SIGNIFICANCE OVER OTHER METHODS FROM PAIRED T-TEST ($P < 0.05$) IS HIGHLIGHTED IN BOLD.

| Method | Liver | | | | Vessel | | RFP(%) | Comput.Time(s) |
|---|---|---|---|---|---|---|---|---|
| | DSC (%) | RAVD (%) | ASSD (mm) | MSSD (mm) | DSC (%) | Disconnection | | |
| No Reg. | 92.85±1.49 | 2.01±1.49 | 3.01±3.59 | 23.31±11.51 | 17.55±0.01 | 1.8±1.48 | - | - |
| Rigid Reg. | 95.99±1.50 | 2.01±1.50 | 1.58±0.59 | 20.26±8.64 | 21.75±0.01 | 1.8±1.48 | 0 | 70.10±17.81 |
| VoxelMorph | 98.39±0.51 | 0.59±0.51 | 0.67±0.14 | 19.09±8.76 | 25.29±0.01 | 1.9±1.25 | 0.01±0.003 | **12.09±4.74** |
| ANTs | 98.59±0.62 | 0.73±0.62 | 0.59±0.09 | 17.60±9.45 | 27.71±0.01 | 2.0±1.45 | **0** | 92.98±25.49 |
| NiftyReg | 98.79±0.61 | 0.85±0.61 | 0.50±0.06 | 17.30±9.42 | 38.33±0.01 | 2.2±1.40 | 0.37±0.28 | 414.10±172.95 |
| Ours | **99.18±0.19** | **0.32±0.19** | **0.34±0.05** | **16.81±9.43** | **1.23±0.01** | 1.8±1.48 | 0.002±0.002 | 15.66±5.00 |

TABLE II
QUANTITATIVE RESULTS OF ABLATION STUDIES IN LIVER DCE-CT DATASET

| Method | Liver | | | | Vessel | | RFP(%) |
|---|---|---|---|---|---|---|---|
| | DSC (%) | RAVD (%) | ASSD (mm) | MSSD (mm) | DSC (%) | Disconnection | |
| U-Net | 98.39±0.51 | 0.59±0.51 | 0.67±0.14 | 19.09±8.76 | 25.29±0.01 | 1.9±1.25 | 0.01±0.003 |
| U-Net+S | 98.81±0.71 | 0.50±0.72 | 0.52±0.14 | 17.73±9.89 | 1.71±0.01 | 2.3±1.57 | 0.007±0.004 |
| U-Net+S+L | 98.88±0.22 | 0.33±0.22 | 0.46±0.07 | 17.34±9.61 | **1.20±0.01** | 2.4±1.71 | 0.007±0.004 |
| U-Net+S+L+V (Ours) | **99.18±0.19** | **0.32±0.19** | **0.34±0.05** | **16.81±9.43** | 1.23±0.01 | **1.8±1.48** | **0.002±0.002** |

Fig. 6 demonstrates visualization results of one representative experiment for various registration methods. In our experiments, we select the image corresponding to arterial phase as the moving image, and the image corresponding to the venous phase as the fixed image for registration. Notice that, the moving image is with a segmentation of liver (indicated by purple in Fig. 6) and a segmentation of arterial vessel (indicated by red in Fig. 6). Correspondingly, the fixed image is with a segmentation of liver (indicated by cyan in Fig. 6) and a segmentation of vein vessel (indicated by green in Fig. 6). As illustrated in Fig. 6, the orange portion (in the second row of Fig. 6) of our proposed method is larger than other methods, which means that our proposed method has better registration performance in terms of liver region. In addition, for unpaired vessel segmentations with complex structures, our proposed method can effectively prevent the overlap between different types of vessels, and preserve the topology structure of registered vessels. For obvious areas, see the corresponding red dashed box area marked in Fig. 6.

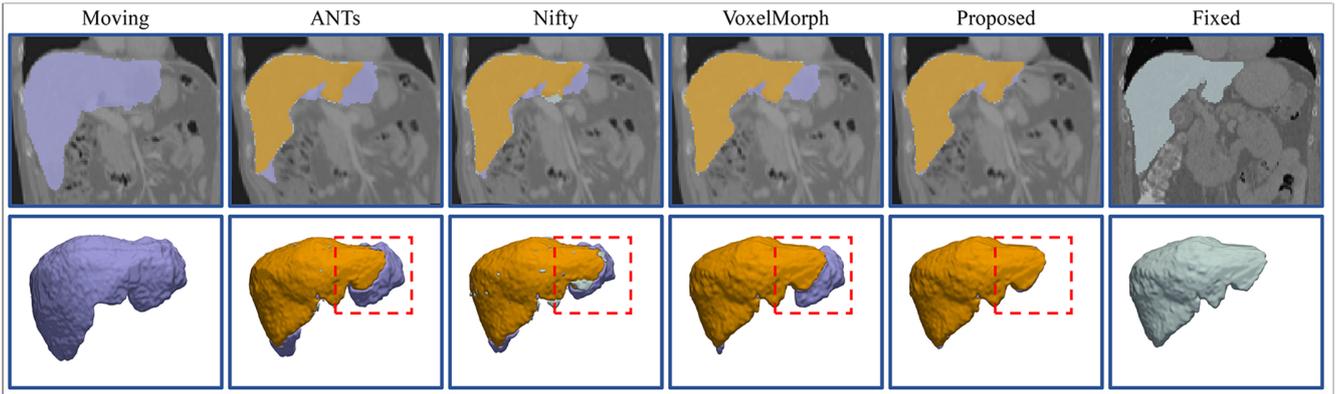

Fig. 7. The registration results of various methods in the LiTS dataset. The orange region represents the overlap between registered liver segmentation and fixed liver segmentation.

*4) Ablation Study:* We further perform ablation studies to verify individual contributions of three components in this work, including segmentation masks $\mathbb{S}$ ($\mathcal{L}_{seg}^l$, $\mathcal{L}_{seg}^v$), structure-aware constraints of liver surface $\mathbb{L}$ ($\mathcal{L}_s$, $\mathcal{L}_{is}$), and structure-aware constraints of vessel centerline $\mathbb{V}$ ($\mathcal{L}_v$, $\mathcal{L}_{iv}$) . Different combinations of these components used in our ablation studies are presented in Table II. For each combination, we use the same training set as described in Section IV-D, and train the networks using the same configuration.

It can be seen from Table II that, without the help of segmentation mask, the evaluation metrics (include DSC, RAVD, ASSD and MSSD) between paired liver segmentation degrade significantly. At the same time, there is a large overlap between arteries and veins after registration. Therefore, the registration accuracy of liver can be effectively improved by using classical DSC loss between paired liver segmentations. Furthermore, the introduction of negative DSC loss (i.e., non-overlap loss) between different types of vessels can also effectively reduce the overlap between them. However, the non- overlap loss between arteries and veins forces vessels within the moving image to look for other similar vessels within the fixed image for alignment, resulting in severe vessel disconnection (especially for some vessels that are spatially adjacent and structurally similar).

In order to solve this problem and further improve the registration accuracy, we incorporate $\mathbb{L}$ and $\mathbb{V}$ in the registration network. As shown in Table II, with the help of $\mathbb{L}$, the registration accuracy of liver can be further improved. Additionally, by combining all components in our proposed method, the registration performance of liver and vessels are improved comprehensively. The above results validate the effectiveness of the proposed three components in the structure-aware network.

### C. Evaluation on LiTS Dataset

*1) Dataset:* Different from DCE-CT dataset, the LiTS dataset contains 131 normal CT scans ( $131 \times 130$ image pairs) with ground-truth liver segmentations. The average resolution of the dataset is $0.80 \times 0.80 \times 1.45 \ mm^2$ , and the average image size is $512 \times 512 \times 448$ voxels. This dataset has been randomly split into 101 ($101 \times 100$ image pairs) and 30 scans ($30 \times 29$ image pairs) for training and testing, respectively.

*2) Pre-Processing:* For the pre-processing of the LiTS dataset, all raw scans are resampled into $180 \times 160 \times 78$ voxels after cropping unnecessary area around the liver region. Then, we mask all the non-liver structures and normalize image intensities to 0-1. Subsequently, we use Ants-Registration to perform rigid alignment of all image pairs.

*3) Comparative Analysis:* Table III lists the registration performances of various methods for the 30 testing subjects in the LiTS dataset. Since non-liver structures are masked during pre-processing, all compared methods can achieve relatively satisfactory registration performance. Specifically, compared with other registration methods, our proposed method has significantly higher DSC and lower RAVD between paired liver segmentations. Meanwhile, the ASSD and MSSD of our proposed method are also significantly smaller than other methods, indicating that our proposed structure-aware network can significantly improve registration performance.

In addition, registration results for the LiTS dataset can also be visually inspected in Fig. 7. Note that the color representation in Fig. 7 is similar to Fig. 6. As can be seen from Fig. 7, there is an obvious difference in the appearance of liver between moving image and fixed image. For such situation, most comparison methods (ANTs, Nifty and VoxelMorph) cannot accurately align liver regions. Compared with other registration methods, our proposed method can achieve better

TABLE III
THE REGISTRATION PERFORMANCE OF VARIOUS METHODS IN THE LITS DCE-CT DATASET, INCLUDING MEAN ± STD DSC, RAVD, ASSD, MSSD OVER LIVER
SEGMENTATIONS, AS WELL AS MEAN RFP FOR DEFORMATION FIELD AND COMPUTATION TIME FOR VARIOUS METHODS.

| Method | Liver | | | | RFP (%) | Comput.Time (s) |
|---|---|---|---|---|---|---|
| | DSC (%) | RAVD (%) | ASSD (mm) | MSSD (mm) | | |
| No Reg. | 46.70±20.74 | 23.65±20.74 | 9.061±3.99 | 32.11±10.45 | - | - |
| Rigid Reg. | 78.96±10.46 | 26.02±10.46 | 3.356±1.41 | 22.44±9.90 | 0 | 20.89±5.81 |
| VoxelMorph | 94.22±0.06 | 7.03±9.84 | 1.21±1.40 | 19.71±10.78 | 0.03±0.02 | **0.79±0.21** |
| ANTs | 93.65±0.04 | 6.86±7.71 | 1.32±0.94 | 16.56±8.68 | **0** | 17.50±2.13 |
| NiftyReg | 94.31±0.03 | 5.81±5.49 | 1.23±0.63 | 14.98±8.43 | 0.02±0.09 | 24.10±2.95 |
| Ours | **97.35±0.01** | **1.48±1.90** | **0.49±0.29** | **12.71±7.98** | 0.05±0.07 | 0.46±0.38 |

TABLE IV
QUANTITATIVE RESULTS OF ABLATION STUDIES IN THE LITS DATASET

| Method | Liver | | | RFP (%) |
|---|---|---|---|---|
| | DSC (%) | RAVD (%) | ASSD (mm) | MSSD (mm) | |
| U-Net | 94.22±0.06 | 7.03±9.84 | 1.21±1.40 | 19.71±10.78 | 0.03±0.02 |
| U-Net+S | 96.87±0.03 | 2.77±4.89 | 0.70±0.81 | 16.48±10.00 | 0.06±0.05 |
| U-Net+S+L (Ours) | 97.35±0.01 | 1.48±1.90 | 0.49±0.29 | 12.71±7.89 | 0.05±0.07 |

registration performance and preserve topology of the registered organs.

*4) Ablation Study:* For LiTS dataset, we also conduct the ablation studies to further verify individual contributions of liver segmentation S and L in our proposed network. Table IV provides performances for different combinations of these components used in our proposed network. Similar to previous experiments, the evaluation metrics of liver is significantly improved with the help of segmentation. Further, the registration accuracy of liver, especially for the surface of liver, can be further improved by the integrating of $\mathbb{L}$ into registration network.

## V. DISCUSSION

In liver DCE-CT image registration, intensity variations caused by the flow of contrast agents, combined with complex spatial motion induced by respiration and heart beats, limit the effectiveness of existing intensity-based registration methods. In this study, we leverage segmentation information to guide the deep learning-based registration network to solve the above problems and comprehensively improve registration performance.

Actually, the segmentation-guided registration strategy has been widely used in various registration tasks, including multi-modality registration [13], [14] and infant brain registration [12]. For instance, during the development of infant brain, the gray matter and white matter tissues exhibit huge intensity differences across the MR images acquired at different ages. Such intensity differences are similar to the intensity variations of liver DCE-CT images, which bring a great challenge for intensity-based registration method. To track this problem, an effective strategy is to employ the segmentation- guided strategy to eliminate the effect of intensity variations. Although the segmentation-guided registration strategy has achieved great success, there are still some limitations to be further addresses. First of all, the existing segmentation- guided registration methods [12]–[14] require paired organ segmentations for guidance, which is not always available in clinical practice. Furthermore, the above-mentioned registration methods [12]–[14] mainly focus on volumetric registration inside the segmentations, and regularize the global deformation field in a uniform manner. These approaches ignore the inherent local structural information of organs (such as topological consistency and smoothness of organ surfaces), and cannot preserve the topology of some fine features effectively.

In order to solve the above problems, we design a structure-aware network by introducing hierarchical structural information to further improve registration performance. Specifically, according to different organ shapes (e.g., tubular or round shapes), we extract structural information from multiple geometrically-meaningful perspectives, and construct additional structure-aware regularization constraints based on the extracted structural information. The extracted structural information is simple-yet-effective to represent complex organ morphology. Since the constructed regularization constraints are directly imposed to the deformation field, our proposed method dose not rely on paired segmentations for guidance. Meanwhile, our designed structure-aware constraint can be regarded as a general regularization module, not limited to particular organs or tissues, can be flexibly applied to networks with various architectures.

However, some limitations still exist in our work. According to the network structure of our proposed method, the number of structural-aware constraint losses depends on the number of organs to be registered. In order to ensure satisfactory registration performance of our proposed method, the weight of each loss in equation (4) need to be manually set, which will cause subjective human errors. It can be known from equations (6) and (9) that the designed structure-aware constraints with the widely-used global regularization term are constructed by derivatives of the displacement field components. Different from equation (5), our designed structure-aware constraint considers more directions. Therefore, the weight of structure-aware constraint loss mainly depends on both the number of voxels within the extracted centerline or surface and the number of directions calculated. In the future, we plan to use a parameter estimation method to select appropriate weights for different losses.

Since the structure-aware constraints are constructed according to extracted structural information, the registration performance of our proposed method still depends on the accuracy of organ segmentations. In the future research, also within the proposed framework, a segmentation network could be integrated to improve segmentation and registration performances simultaneously.

## VI. Conclusion

A structure-aware deep learning framework for liver DCE-CT registration is proposed in this paper. The intensity variations and complex spatial motion within the liver are two major challenges in liver DCE-CT image registration. To address this problem, we leverage structural information of different organs to supervise the training of segmentation-guided deep registration network. Specifically, we extract structural information from different organs, and then construct structure-aware constraints based on the extracted information. Subsequently, the structure-aware constraints are regarded as regularization term, and imposed on forward and backward deformation fields simultaneously. In this way, our proposed network can *not only* handle the intensity variations and preserve organ topology during registration, *but also* utilize unpaired organ segmentations for registration. The experimental results on the testing data indicate that the proposed method can achieve better registration accuracy compared with state-of-the-art registration methods.